# How do Practitioners Perceive the Relevance of Requirements Engineering Research? An Ongoing Study

Xavier Franch[1], Daniel Méndez Fernández[2], Marc Oriol[1], Andreas Vogelsang[3], Rogardt Heldal[4], Eric Knauss[4], Guilherme Horta Travassos[5], Jeffrey C. Carver[6], Oscar Dieste[7], Thomas Zimmermann[8]

[1]Universitat Politècnica de Catalunya (UPC), Spain. {franch, moriol}@essi.upc.edu
[2]Technical University of Munich, Germany. daniel.mendez@tum.de
[3]Berlin Institute of Technology (TU Berlin), Germany. andreas.vogelsang@tu-berlin.de
[4]Chalmers | University of Gothenburg, Sweden. heldal@chalmers.se, eric.knauss@cse.gu.se
[5]Federal University of Rio de Janeiro/COPPE and CNPq Researcher, Brazil. ght@cos.ufrj.br
[6]University of Alabama, United States. carver@cs.ua.edu
[7]Universidad Politécnica de Madrid, Spain. odieste@fi.upm.es
[8]Microsoft Research, United States. tzimmer@microsoft.com

*Abstract—* The relevance of Requirements Engineering (RE) research to practitioners is a prerequisite for problem-driven research in the area and key for a long-term dissemination of research results to everyday practice. To better understand how industry practitioners perceive the practical relevance of RE research, we have initiated the RE-Pract project, an international collaboration conducting an empirical study. This project opts for replication of a survey previously conducted in two domains. We have designed a survey to be sent to ask industrial practitioners to rate their perceived practical relevance of 418 RE papers, published between 2010 and 2015 at the RE, ICSE, FSE, ESEC/FSE, ESEM and REFSQ conferences. We plan to send this survey to several hundred industry practitioners at various companies around the world. In this paper, we summarize our research protocol, present the current status of our study, and describe the planned future steps.

## I. INTRODUCTION

High-quality Requirements Engineering (RE) directly contributes to appropriateness and cost-effectiveness in the development of a system [6] whereby RE is a determinant of productivity and (product) quality [7]. Yet, RE remains inherently complex due to the various influences in industrial environments. This complexity makes the choice of adequate processes, methods, and tools dependent on the needs and particularities of the practical contexts as in no other software engineering discipline. This dependence makes it impossible to standardize RE via holistic and universal solutions.

Over the last years, we have observed an active research community arise and propose a plethora of promising contributions to RE. However, we still know very little about the practical impact of those contributions or whether they are in tune with the practical problems they intend to address [8]. In fact, there still seems to be often a gap between research and current practice [3]. It was, to our knowledge, first discussed in 2000 at panels during the 12th International Conference on Advanced Information Systems Engineering (CAiSE) and the 4th International Conference on Requirements Engineering (ICRE), and then later summarized by Kaindl et al. [9]. Recent panels at the International Requirements Engineering Conference on obstacles for technology transfer into practice as well as ongoing debates (as recent as in the last edition of the Working Conference on Requirements Engineering: Foundations for Software Quality –REFSQ 2017, following the keynote by Lionel Briand) on the extent to which RE research and practice are detached from each other highlight the need for a radical change in the community [10]. This apparent need raises the following questions: (1) Do practitioners perceive academic RE research to be relevant to their work and (2) How can scholars make RE research (even more) relevant to practitioners?

Motivated by a similar line of thoughts, Lo et al. [1] performed a study to assess how practitioners at Microsoft perceive the relevance of software engineering papers published at ICSE, ESEC/FSE and FSE from 2009-2014. Carver et al. then replicated this study to gauge how a broader sample of practitioners perceived the relevance of ESEM papers published between 2011 and 2015. In this joint work, we now plan to conduct a second replication for the RE community to understand whether RE research in is disconnected from practitioners' needs.

In this paper, we summarize our research protocol and present the status of our study and the planned future steps. The rest of the paper is organized as follows. In Section II, we state the objective of the study and then elaborate our research questions. In Section III, we introduce the context of the study. In Section IV, we elaborate on the study plan and discuss the threats to validity in Section V, before concluding the paper in Section VI.

## II. OBJECTIVE OF THE STUDY

The primary goal of the RE-Pract project is to investigate the RE practitioners' overall perception of the practical relevance of currently published RE research. To achieve this goal, we define five research questions (RQs). The first four ones are consistent with the previous surveys [1][2], which form the basis for our replication. The final RQ emerges from the particularities of RE as an interconnected discipline.

The first RQ forms the central focus of the work to understand the general perception of the practical relevance as perceived by practitioners from industry.

**RQ1: What is the relevance of RE research to practitioners in industry?**

This first RQ builds the core of our investigation, yet we naturally aim at gathering further details that provide a broader and more detailed picture of practitioners' perceptions. To this end, we add further RQs. We expect that the practitioners' perception of the importance of research is also influenced by the topics addressed rather than based on the particularities of the individual papers only. Therefore, the next RQ is:

**RQ2: Which research ideas do practitioners rate most highly?**

As our assumption is that there is often a gap between the focus of academic research and the needs of practitioners, the third research question seeks to bridge this gap:

**RQ3: Which research problems do practitioners think are most important for the RE community to address?**

Next, we are interested to know whether papers with direct links to industry influence this perception. These links may manifest as either (i) one or more authors of the paper have an industrial affiliation (often through the research arm of an organization) or (ii) the paper appears in the industry track (for conferences that contain such a track):

**RQ4: Do papers with explicit ties to industry have higher practical relevance than other papers?**

Finally, to obtain more insight into the overall perception of relevance, we would like to understand whether the respondent's role in his company is a factor. We believe this factor is important in RE because the discipline (and its outcome) is affected by various, potentially differing, needs and expectations among the stakeholders involved. Thus, want to know:

**RQ5: Do practitioners' perceptions and views differ depending upon their roles?**

## III. CONTEXT OF THE STUDY

### A. Background

The RE-Pract project is a joint initiative of the first two authors of this paper (Franch and Méndez Fernández) after attending a keynote given by the last author (Zimmermann) at the 9th ACM/IEEE International Symposium on Empirical Software Engineering and Measurement (ESEM 2015) held in Sept. 2015 in Beijing, China. As part of the keynote, Zimmermann reported the study published in ESEC/FSE 2015 [1] and conducted by Lo (Singapore Management University), Nagappan and himself (Microsoft Research) to understand practitioners' perception of the relevance of software engineering research in general. The study was based on rating a random sample of the research published in 571 papers at the ICSE, ESEC/FSE, and FSE conferences in the period of 2010-2014. Overall, they gathered 17,913 ratings by 512 practitioners from Microsoft. The findings were organized around three research questions: 1) how do (Microsoft) practitioners view software engineering research as a whole?, 2) what research ideas do (Microsoft) practitioners consider to be most important? and 3) why do (Microsoft) practitioners view some research ideas as unwise? A high number of the respondents, 71%, provided overall positive ratings.

This ESEM 2015 keynote resulted in a lot of interest from the audience and formed the seed for at least two independent follow-up studies. Carver et al. conducted the first study [2] focused on the empirical software engineering community. The main drivers of the replication were two of the authors of the current paper (Carver and Dieste) in collaboration with a third author from industry (Kraft from ABB Corporate Research) and two authors of the initial study (Lo and Zimmermann). The resulting study (published in ESEM 2016) gathered 9,941 ratings by 437 practitioners of 156 papers published at the ESEM conference between 2011 and 2015. The overall percentage of positive ratings was close to the former one, namely 67%.

The second replication is this current paper. Shortly after ESEM 2015, the first configuration of the team (seven first authors plus last author of this paper) was completed and started working. After becoming aware of the first replication [2], its first two authors, Carver and Dieste, were invited to join the team, leading to the final team of authors.

### B. Issues, Pitfalls, and Mitigations

The first two authors have previously initiated international collaborations around RE topics involving contributors from multiple countries. The first is NFR4MDD (Non-Functional Requirements for Model-Driven Development[1] [4]), initiated by the first author to investigate the adoption of non-functional requirements in the context of model-driven development in industrial settings. The second is NaPiRE (Naming the Pain in Requirements Engineering[2]), initiated by the second author forming a collaboration with currently more than 50 researchers worldwide and including a bi-yearly replicated family of distributed surveys investigating the current state of RE practices and problems encountered therein. Both projects are different in the topics addressed and the research methods applied but are comparable to each other and to the study at hand from the perspective of potential issues and pitfalls. In addition, the inclusion in the team of authors of the former baseline studies [1][2] should help to anticipate and mitigate possible barriers. Even so, we remain aware that we may very well expect further ones to arise in later stages of the study execution, e.g. coming with changes of affiliations.

As can be expected from a paper by authors from eight organizations, one first basic issue concerns the overall coordination and decision-making. In our case, the first two authors proposing this initiative are managing the organizational set-up of the project and related organizational tasks (e.g. time schedule proposals, coordination of the communication, or proposals of workload distributions). The first two authors also coordinate any decisions on issues related to the scope and design of the project itself, but make the final decisions jointly with the other authors.

---

[1] http://www.essi.upc.edu/~gessi/NFR4MDD/index.html

[2] http://www.re-survey.org

Another basic but important issue concerns the establishment of a commonly shared infrastructure. Here, we made a pragmatic decision and set up a shared space in Google Drive to support collaborative editing of documents with good traceability features and to share the several required documents (previous studies, study protocols, etc.).

Team members' communication is another challenge in such a project. The time difference between the easternmost and westernmost partners is nine hours, rendering it difficult to set up live team meetings. We created a mailing list as the main communication channel in the project. Together with the shared space, this simple yet effective solution is the primary team communication channel.

Author order is another common issue for larger research teams. We jointly decided in advance to make decisions on a case-by-case basis following the classification of contributing roles for authorship as proposed by Brand et al. [12] and previously adopted in the context of NaPiRE [11]. For this paper, we had three main categories of authors: main contributors driving the overall project and building the core team for the writing, members involved in the preparation of the data collection, i.e. creators of paper summaries for this study (see also the next section), and advisors with experience in the two previous studies and their design. We sorted each category alphabetically. Besides, each paper will describe the responsibilities and work undertaken by each author.

Finally, as mentioned above, further issues might arise during the project execution. At the end of the next section, after introducing the overall study design (planning), we briefly discuss current open issues.

## IV. PLANNING OF THE STUDY

The overall goal of the RE-Pract project is to investigate practitioners' perceptions of the practical relevance of today's academic research in Requirements Engineering. Structuring it more precisely and following the Goal Definition Template [5], we want to

**Analyze** RE academic contributions (research ideas, tools, approaches, methods, and techniques)
**in order to** characterize
**with respect to** the perceived practical relevance
**from the point of view** of Software Engineering practitioners (requirements engineers, architects, testers, etc.) dealing with requirements
**in the context of** full (published) research papers

The subjective views of practitioners on academic research outcomes are dominated by their everyday practice, experiences, beliefs, and personal taste. Therefore, we design our project as a qualitative study relying on survey research. To address a broad population, we opt for online survey research designed as an anonymous survey to lower potential barriers to participation.

The main audience of this research is practitioners working with requirements in industrial settings in one form or the other (ranging from requirements engineers to testers). Their key motivation to participate in the study is, similarl as in the NaPiRE project, their contribution to increasing the awareness of topics they considered important. The main audience of our research outcomes is the overall RE research community. Our hope is that the results support ongoing reflections on the practical relevance chosen research topics might have (without any prejudice to the individual judgment of the researcher herself and without judgment about papers where the practical relevance is not and should be not the primary quality attribute).

In the following, we briefly introduce the overall study planning including:
1. Paper selection and summarization.
2. Participant selection.
3. Feedback elicitation via survey to gather ratings of research summaries from Step 1.
4. Data analysis with respect to the research questions.

In the following, we elaborate details while focusing on the first three items.

### A. Paper Selection and Summarisation

The first data collection step is the selection and preparation of the papers to be rated by practitioners. To this end, we extracted a pool of 418 papers published between 2010 and 2015 at the RE, ICSE, ESEC/FSE (including FSE when held alone), ESEM and REFSQ conferences. Because we are aware that early stage solution proposals, such as visionary papers, might not attract the interest of practitioners despite their potential value in the future, we intentionally decided to concentrate on full papers only to prevent distortion of the results. We included all full papers from the research and industry tracks, even if, for some conferences, industry track papers are required to be shorter compared to research track papers. We thus excluded short, vision, or ongoing research papers regardless of research or industry track.

For each paper, we created a short, one sentence summary of its scope. In contrast to the baseline studies [1][2] where the authors of the selected papers provided the paper summaries themselves, we used the original abstracts (and in cases of doubts the paper's body) and created the summaries on our own. The main reason for writing our own summaries was to ensure consistency among the summaries, to reduce the effects of an author's ability to write an appealing summary, and pragmatically we deemed it impossible to contact all involved authors given the broad spectrum of venues involved. We created our summaries in pairs of researchers. After the summary creation, another pair of researchers then validated the overall outcome.

Each summary included the main contribution of the paper and the potential research type facets [13], such as "solution proposals" or "evaluation". For instance, for a paper proposing and evaluating a specific requirements elicitation technique, we formulated the summary in the form "An evaluated requirements elicitation technique that [details of the technique]." We crafted the summaries for RE and REFSQ from scratch. For the papers published at ICSE, ESEC/FSE, FSE and ESEM (which were already summarized in the previous studies), we excluded papers not related to RE and added any missing RE papers published from 2010 to 2015. We merged the summaries into one spreadsheet and revised the summaries to fit the intended structure. Finally, for each paper, we documented (in addition to the authors' names and abstracts), the venue, the year, whether the authors had any ties to industry based on their affiliation (*academic* in case all authors were from academic institutions, *industry* in case all authors were from industry, or *mixed*), and whether it was an industry track submission or not.

### B. Participant Selection

We chose individual practitioners as the unit of analysis. Those practitioners need to have clear ties to RE in their everyday practice, i.e. their roles and responsibilities include both creating and managing requirements, or working with them in a broader sense (e.g. architects or testers). To select the participants, each of the authors created a list of personal contacts to industry. We followed the same strategy as in the NaPiRE project and opted for an invitation-based survey where we approach individually known practitioners rather than distribution the survey randomly based on, for instance, mailing lists or social media channels for mainly two reasons. First, relying on a list of known contacts helps ensure that the respondents have the necessary background to provide useful answers. Second, inviting known respondents gives us the chance to control the responses and the response rates. Even if the responses remain anonymous, to reduce barriers that might hinder respondents to reveal their real opinions, we believe that the invitation of known practitioners helps us equally distribute the survey among various companies. That is, we are interested in the views of the individuals, independent of company-specific views. Therefore, we choose to distribute the survey equally among various practitioners from multiple companies to reduce the risk of having too many practitioners from a single company. The downside of this approach, of which we are very cognizant, is that it might yield lower numbers of participants than in the previous studies [1][2].

### C. Feedback Elicitation

Following the design of the baseline studies [1][2], we plan to use an online survey. We will design the survey such that it requires as little effort as possible for participants to complete it. For example, the survey will be self-contained and will include all relevant information. We will limit the response types to numerical, Likert-scale, and short free-form answers. As part of the questionnaire, we will elicit feedback in three categories while staying as close as possible to the questions in the baseline studies:

**Demographics:** Collecting this basic information about the participants allows us to break down the results by, e.g., roles (such as developers or testers) or domains.

**Ratings of research ideas**: We will present a subset of randomly selected paper summaries to each participant (in a random order). For each summary, the respondent must rate the research idea based upon the question "In your opinion, how important are the following pieces of research?". We will use the same rating categories as used in the baseline studies.. Participants can label a research idea as "Essential," "Worthwhile," "Unimportant," "Unwise," or "I Don't Understand." The last category is included to address the diverse background of participants—not all participants will understand all technologies.

**Qualitative Feedback:** We will additionally ask for two types of qualitative feedback. First, to understand the rationale behind the ratings, we will randomly select two of the summaries the participant rated and ask them to "provide a brief explanation for why you found it either relevant or not to your work." Second, we will give the participants an opportunity to provide guidance to the research community about topics of interest. We will ask them "Suppose that you could provide guidance to a team of RE researchers, what problems should they focus on first?".

### D. Current Stage

At the time of paper writing, we have completed the selection and summarization. In the following Table, we illustrate the distribution of the final pool of papers and the ratio of papers with ties to industry (i.e. papers with at least one industrial co-author and / or industry track papers). Not surprisingly, the RE and REFSQ conferences greatly dominate the distribution due to their focus on RE.

| Venues | Number of papers | Industry ratio |
|---|---|---|
| RE | 212 | 32,5% |
| REFSQ | 144 | 18% |
| ICSE & ESEC/ FSE | 43 | 23% |
| ESEM | 19 | 52% |

We will publish a detailed summary of the papers once we have completed the data analysis. Currently, we are finalizing the list of individual contacts from industry and the questionnaire. Once we have completed the questionnaire, we will implement it as a web application using the *Enterprise Feedback Suite* and pilot it with practitioners.

### E. Open Issues

In these initial steps of study design, we face some issues that have prompted interesting discussions. Most prominently, we are discussing the concept of *industry papers, which* comes with a non-trivial question: When does a paper qualify as an industry paper? So far, we rely on a (not mutually exclusive) classification via authorship and the track in which papers appear. In the case of authorship, we classified whether at least one or all authors of papers have industry affiliations, i.e.

affiliations to companies or related research units. This, however, is itself a non-trivial decision as many researchers have nowadays multiple affiliations, e.g. researchers working at both a university and a company. Further, the notion of "industry" is fuzzy itself as it is not often clear how to classify institutions that perform research and transfer to bridge the gap between classic companies and, for example, universities (such as Fraunhofer institutes). Currently our plan is to collect as much data as possible about the papers as separate data items so that we can leave the options open and decide later how best to aggregate the information.

Another open question, that came from discussions during the baseline studies but has not yet been realized in a survey, is that after practitioners provide feedback, we would like to provide them with pointers to the papers describing the research ideas they rated as highly relevant. This additional step could help to strengthen the ties between academic researchers and industry participants.

Finally, another open issue regards the population source for the industry contacts. We deliberately decided to rely on personal contacts only and not to spread the survey invitation anonymously using available channels (e.g. mailing lists). It also means not to include the contacts at Microsoft as this potentially high number of responses from one company alone might be in strong contrast to the otherwise diverse but smaller number of responses from our contacts lists. However, we are still discussing how to increase the population sample within the limits of our existing constraints.

## V. THREATS TO VALIDITY

As any other empirical study, this project is facing some threats to validity. Some of them are already known, while others may appear later as the study progresses. We briefly report them together with associated mitigation actions, relying on the classification as proposed by Wohlin et al. [5].

### A. Internal Validity

**Abstract comprehension**. We intentionally decided against using the original, longer abstract published with the paper in favour of formulating our own short summaries. The primary motivations for this choice were (1) avoiding the risk that long summaries would results in a high mortality rate for survey respondents and (2) minimizing the role that abstracts designed to communicate with researchers rather than practitioners could have on practitioners' perception of the research ideas. To mitigate this risk, we very precisely defined what a summary must describe. We included a validation step to harmonize the summaries that different team members created.

### B. External Validity

**Representativeness of papers**. We selected a set of venues and a set of years (2010-2015) from which to draw papers for this study. We did not include papers from 2016 because we started the work in the middle of 2016. We chose the venues by selecting those that are considered to be world-leading conferences related to RE (RE and REFSQ) and to software engineering in general (ICSE, ESEC/FSE, FSE, and ESEM). Even with this choice, we realize that including more (or different) venues could affect the study results. In addition, given that conference papers tend to have a smaller and more digestible focus compared with the broader research described in journal papers, we intentionally concentrated on research published in conferences and excluded research published in journals.

**Representativeness of respondents**. Even though at the moment we do not yet know how many responses we will obtain or the extent to which some companies could be dominant (as it happened with Microsoft in the two previous studies [1][2]), we definitely need to consider this threat. To mitigate it, we are planning to use as many additional practitioner networks as possible. In the field of RE, we have some resources that worth to consider: the NaPiRE database, the IREB magazine which is well-known by European RE practitioners, but over and above all our own networks (and some of the paper authors have ample networks), we plan to involve some RE practitioners who may be especially sensible to this issue (e.g., because they are usual attendees of the RE conference).

### C. Conclusion Validity

**The meaning of "perception."** This study focuses on how industrial practitioners *perceive* the relevance published academic research papers. Note that it is not our intention nor do we pretend to over claim the observations gathered in the study. For instance, we will not claim that highly-ranked papers or research areas are more likely than others to be adopted by practitioners or that they will have a higher impact than others. We are definitely aware that a relevant problem may not be addressed in a relevant way. In fact, we are very much aware that the practical relevance of research can truly be judged only after the fact based on the extent to which the ideas have been adopted or not. However, our position is that the results of the study can provide a good first indicator of such impact. Ultimately, these results serve to foster discussions on important aspects in our field given the practical scope many contributions have.

**Replicability**. One major prerequisite for replicability (and more generally reproducibility) is the openness of the study design as well as the data obtained. Therefore, once we have finalized the study, the protocol and all the related material used to perform this study, we will make those materials available under CC-BY license. Furthermore, we plan to publish all study data in an open repository for other researchers to use. The open character of our project will help researchers and practitioners replicate this study and, in the long run, help to better generalize further from the results.

### D. Construct Validity

**Methodology robustness**. The robustness of a methodology depends on many (non-trivial) factors, and many threats may arise during the planning phase. For instance, the protocol may be incomplete, it may lack necessary details, or it may even contain flaws with respect to the data analysis. Furthermore, the research questions may be incomplete or the questionnaire insufficient to answer the research questions. As a mitigation to this threat, we stay as close as possible to the original studies upon which this replication study builds. Furthermore, the author team, many of whom worked on prior studies, has discussed the research protocol. For the sake of transparency and as a means of quality control, we will further disclose all the data including the detailed protocol. We finally include a validation phase for the survey, which will include a pilot of the analysis methods planned. This pilot phase will allow us to, at least, control and tackle the most severe issues in advance and apply \corrective measures before dissemination of the survey.

## VI. CONCLUSIONS

In this paper, we have reported on the planning and current status of an ongoing empirical study to gauge the perception of practitioners about published academic research in RE. This study may benefit several stakeholders: for researchers looking for transferability of their results, it may suggest areas of future research (or fine-tune the ones currently in their scope of investigation); for practitioners, it may help them discover lines of current research that could eventually be interesting to them, and it provides them with the ability to add their own views and flavours; for conference organisers, it may help to assess the topics that they offer to the community.

In the context of our study, this paper represents a milestone. First, we aligned the deadline with the finalization of an important activity, namely the completion of the paper summaries. The paper became a motivating instrument to speed up in the finalization of this activity. Next, the organization of the team has improved given the need of collaborating to finalize this paper on time and in some sense, it has become a proof of concept for the way of working for the rest of the study. Last, we acquired relevant feedback from the paper reviewers for improving our protocol promptly. Once at the conference, the presentation itself will provide an excellent opportunity to share our first impressions on the ongoing analysis and raise awareness of the study.

As mentioned already, this is the third study of its kind with a very similar protocol. We plan to compare our results with the two prior studies and start to identify trends. Since the protocol will be disclosed to the public, we cordially encourage other researchers to replicate it for other areas (software architecture, testing, etc.) to get a deeper understanding of the (perceived) practical relevance of software engineering in general.

## ACKNOWLEDGMENTS

This study has been partially funded by the project TIN2016-79269-R.